# Credibility of Automatic Appraisal of Domain Names

Karol Król
University of Agriculture in Kraków, Department of Land Management and Landscape Architecture, Al. Mickiewicza 24/28, 30-059 Kraków, Poland
Email: k.krol@urk.edu.pl

Artur Strzelecki
University of Economics in Katowice, Faculty of Informatics and Communication, Ul. 1 Maja 50, 40-287 Katowice, Poland
Email: artur.strzelecki@ue.katowice.pl

Dariusz Zdonek
Silesian University of Technology, Faculty of Organization and Management, Ul. Roosevelta 26, Budynek A, 41-800 Zabrze, Poland
Email: k.krol@urk.edu.pl

*Abstract*—Both domain names and entire websites are increasingly frequently treated as assets, the value of which can be appraised. The objective of the present thesis was to verify the credibility of domain name appraisals obtained using generally available web applications in an automated, algorithmic way. In conclusions section, it was mentioned that the terms "domain name appraisal" and "website appraisal" are frequently equated. It was also shown that algorithms used in the tested applications consider parameters characterising websites. Thus, they cannot be used to verify the value of domain names themselves. Moreover, during the analysis of the pattern of operation of the appraisal websites it was noticed that they were not made available with domain name or website appraisals in mind. Their objective was to acquire and intercept online traffic. Such applications also left cookie files on recipients' devices, which were then used by advertising systems based on the re-marketing concept.

## I. Introduction

INTERNET became an irreplaceable means of communication, source of information and entertainment. It is also increasingly frequently used to conduct economic activity [1]. It enables a quick and massive flow of information. It is characterised by interactivity and social potential, which allows users to exchange opinions about products and services. Internet offers easy and relatively cheap communication with clients, measurability of undertaken business activities and it enables their modification on a current basis [2], [3].

The economic sector based on online activities mainly consists in commercial use of websites. A website and the related URL frequently constitute a highly valuable capital of an enterprise. Moreover, as internet's popularity grew, its users noticed that a web address can bring diverse profits. This stimulated the development of a domain name market, among others, which is currently one of the most dynamically developing markets around the world [4], [5]. One of the reasons for the growing popularity of domain names is the fact that websites have become one of the most efficient marketing tools [6]. Domain names can be registered on the primary market or one can purchase rights to a domain name on the basis of a contract for registration of a domain name on the secondary market. By registering a domain name, the subscriber acquires the right to use it or resell it. The subscriber can also extend the binding term of subscription contracts.

The domain name market is distinctive. On the one hand, there are known criteria, on the basis of which one can determine the value of a particular domain name, while on the other hand, the value of concluded transactions and the price of domain names offered on the secondary market are very diverse [7].

A purchase of a domain name on the secondary market frequently entails difficulty regarding the assessment of its value. On numerous occasions, it is also connected with the performance of an appraisal of the website or web application associated with the domain name. The objective of the present thesis was to verify the credibility of domain name appraisals obtained by using generally available web applications in an automated and algorithmic manner.

## II. Materials and methodology

The first step was to select 10 offers of domain names for sale on the Polish market with a large price spread which were offered through nazwa.pl's online auctions on the Domain Name Market. Secondly, their value indicated in relevant auctions was recorded. Then, an attempt was made to appraise them using selected English and Polish web applications (Table 1).

TABLE I.
DEFINING CHARACTERISTICS OF FIVE EARLY DIGITAL COMPUTERS

| No. | English website appraisal web application | Polish website appraisal web application |
|---|---|---|
| 1 | siteprice.org | wycena-www.pl |
| 2 | yourwebsitevalue.com | wycenastron.pl |
| 3 | siteworthtraffic.com | speedtest.pl/wycena |
| 4 | howmuchismysite.com | wycen-strone.pl |
| 5 | checkwebsiteprice.com | – |

Nazwa.pl is one of the greatest Polish domain name registrars and suppliers of online services. The Domain

This work was not supported by any organization

Name Market service is a platform for communication which allows users to publish and negotiate domain name sales offers. To post an offer on the Market one needs to be a registered user (authorised to access the client's panel). Thus, sales offers can be posted solely by entities authorised to dispose of the domain name.

### III. DOMAIN NAME APPRAISAL OR WEBSITE APPRAISAL?

Both domain names and websites are increasingly frequently treated as assets, the value of which can be appraised. This is because there is a possibility to include domain name's value among company's intangible assets. A reliable and credible appraisal of a website is not an easy task and it involves selecting criteria which determine the value of a website. In most cases website appraisal relies on the broadly understood quality of the website. It affects its perception by users, which can directly translate into conversion of objectives and, consequently, into the value of the website [8]-[10].

Factors which affect the value of a website, which are most commonly mentioned with regard to its appraisal, include: the website's history, synthetic domain name authority indices, website users' profile and their loyalty, market trends, cost of maintenance and development of the website, website's age, number and quality of links leading to the website, its share in social media, functionality, number of visits and unique users, potential cost of creating an application offering a similar functionality, visibility of the website in search results, brand as well as market and industry's characteristic features [11]. Moreover, website's technical attributes are strictly connected with each other and the low condition of any of them can directly affect the others.

Depending on the appraisal concept, the list of criteria which are taken into consideration can vary [12]. It is difficult or even impossible to align the impact of all criteria on the appraisal on the global market. Despite that, there are new criteria, methodologies and algorithms appearing, which are aimed at making domain name and website appraisal credible, which is similar to the case of real property appraisal.

Website appraisal resembles the appraisal of a developed property to some extent. Buildings (residential, utility) can be modernised or expanded, similarly to a website. A plot of land is just as unique as a domain name. In case a real property is used for commercial purposes, its appraisal should involve an appraisal of value resulting from its position on the market (e.g. the number of clients, market share, competitors' characteristics, ability to generate profits, opinions of potential clients). Its value can be determined as an "income value", since it is directly connected with potential income of the property's owner. The appraisal of that component can prove to be the most difficult one. If one were to assume that an appraisal of a real property used for commercial purposes can be equated with a website appraisal, it can be determined that its value consists of three primary components – domain name's value, application's value and income value. This is the point at which website appraisal should be distinguished from domain name appraisal. Website appraisal is different than domain name appraisal, in particular in terms of appraisal criteria (Table 2). This is because the value of a domain name depends on factors such as: the length and popularity of words used in the name, its suffix, characteristic features of the industry and market on which it can be used and transaction prices regarding the sale of similar domain names on the market [7].

TABLE II.
SELECTED WEBSITE APPRAISAL CRITERIA

| Domain name value (selected domain name appraisal criteria):<br>– name's length;<br>– popularity (commonness) of words used in the name;<br>– domain name's suffix;<br>– characteristics of the industry and the market on which the particular domain name can be used;<br>– transaction prices of purchases and sales of similar domain names and websites on the market. |
|---|
| Web application's value:<br>– cost of creation of an application offering similar features. |
| Income value:<br>– number of unique users on the website;<br>– number of website visits;<br>– number of registered users (clients);<br>– competitive position on the market;<br>– website's visibility in online search engines;<br>– type of the market, industry and trends. |

The general value of a website includes the value of the domain name and web application as well as potential for generating income based on its market position. The value of an application can be determined on the basis of potential costs of creation of a website offering similar functionalities. It may be difficult to determine the income value of a website, since it involves numerous components, such as these connected with: website's position in search results, characteristics of the industry, market share, competition and the number of unique users.

The simplest way to appraise websites is an automated, algorithmic analysis using one of the many tools which are available for free. The way individual tools work can be different and the appraisal can be performed based on (estimated) data, such as the number of visits to the website, the value of synthetic domain authority value indices, etc. Such appraisal is usually free-of-charge, but its result is uncertain. An appraisal performed by an expert is an alternative solution which usually includes a detailed description of methodology behind the appraisal.

### IV. SOURCES OF DATA FOR AUTOMATIC VALUATION SYSTEMS

Sources of data for automatic appraisal systems can be primary sources in the form of data collected by the system and secondary sources, i.e. those acquired from diverse sources. Secondary sources include, for instance, Wayback

Machine (Internet Archive) website resources which makes archived copies of websites from around the world available, as well as the AlexaRank ranking, which is a synthetic domain name value index estimated on the basis of the number of visits and time spent by users on a particular website. Link exploration systems can be considered as another source of secondary data. They include Ahrefs, Majestic, MOZ and SEMrush. They automatically collect information about links (connections between websites) available on the internet. Then, such collected data is made available to users of these systems. Moreover, secondary data can be obtained from systems analysing the visibility of websites in search results, e.g. SEMSTORM, SENUTO, Searchmetrics and SimilarWeb. Such tools collect full information about website's visibility in search results of the most popular search engines around the world.

## V. RESULTS AND CONCLUSIONS

The tested web applications perform website appraisals. Their algorithms consider parameters characteristic to websites. Despite that, in many cases their creators assure their users that the applications perform domain name appraisals – the terms are frequently equated. However, appraisal applications cannot be used to verify the value of domain names as such.

Appraisals performed by web applications are radically different from the amounts indicated on the domain name market and they are most frequently dependent on whether a website is connected to a particular domain name (Table 3). This is because the tested web applications are based on algorithms which evaluate the number of unique users and website views (daily, monthly and annual statistical data), the value of synthetic domain name authority indices, AlexaRank statistics, presence in social media and many other factors.

Algorithms used by applications performing tests omit aspects which can be assessed only by a user, such as attractiveness, potential, memorability, recognition and name's length, how easy it is to type it, pronunciation (sound and syllabification), associations and the presence of similar names. Therefore, if a particular domain name is not tied to a website which operates on the basis of the value of aforementioned indices, its value is low or equal to zero according to the appraisal algorithm. Thus, it is possible that a domain name such as "to-pw45xvctzro.pl" would be appraised as one of higher value than a generic one, the pronunciation of which is simpler, but with no website tied to it, i.e. if it was registered only for the purpose of reselling it (there is no established position in search results, no history or statistical data on website use). This is confirmed by results of the analysis. Two of the 10 appraised domain names were tied to a website and they were the ones to be appraised at the highest value according to valuation algorithms. In light of aforementioned, there is a question about how values such as website views statistics are read? It is common for them to be unavailable to the general public and they frequently constitute a trade secret. Values read by appraisal algorithms can thus be only approximate, estimated.

## VI. SUMMARY

Domain name appraisals obtained automatically do not correspond in any way with the amounts indicated as their sale price in online auctions. Therefore, such appraisals should be approached with a healthy dose of scepticism. Paradoxically, is should be concluded that the appraisal applications serve their purpose. This is because by analysing principles behind their operations it can be noticed, that they were not made available with actual domain name appraisal or website appraisal in mind. They are mainly a component of the so-called diagnostic and informational portal infrastructure and their income is derived from advertising. Services they provide only serve the achievement of their goal, while their quality is of secondary importance. Websites appraising the value of websites usually constitute characteristic "environment" for principal websites and they perform certain tasks for their creators. Their objective is "traffic concentration" – acquisition and interception of online traffic. They are a kind of a seizing funnel. Through numerous services provided

TABLE III.
WEBSITE APPRAISAL PERFORMED USING POLISH (AMOUNTS IN PLN) AND ENGLISH WEB APPLICATIONS (AMOUNTS IN USD)

| No. | Web address | Price indicated at the domain name market | | Appraising portal | | | | | | | | |
|---|---|---|---|---|---|---|---|---|---|---|---|---|
| | | PL* | USD** | 1 | 2 | 3 | 4 | 5 | 6 | 7 | 8 | 9 |
| 1^ | doradcapodatkowy.com.pl | 200,000.00 | 49,072.53 | 1476 | 17 | 0 | 5.20 | 93 | 3900 | 0 | 0.5 | 77 |
| 2^ | korepetycje.pl | 120,000.00 | 29,443.52 | 5350 | 43 | 2687 | 5.57 | 2699 | 1628 | 293 | 145 | 3314 |
| 3 | showhouse.pl | 50,000.00 | 12,268.13 | 1872 | n/a | 0 | n/a | 65 | 2899 | n/a | 0 | n/a |
| 4 | styropian-grafenowy.pl | 15,000.00 | 3,680.44 | 2355 | 15 | 0 | 0 | 160 | 92 | 0 | 0 | 1 |
| 5 | szybkikurczak.pl | 10,000.00 | 2,453.63 | 1980 | 15 | 0 | 129.6 | 66 | 147 | 0 | 0.03 | 85 |
| 6 | szybkie-zlecenia.pl | 3,900.00 | 956.91 | 1548 | 15 | 0 | 0 | 155 | 218 | 0 | 0 | 1 |
| 7 | mvw.pl | 2,100.00 | 515.26 | 1556 | 15 | 0 | 0 | 67 | 2040 | 0 | 0.03 | 44 |
| 8 | flooring3d.com | 100.00 | 24.54 | 2045 | 15 | 0 | 0 | 51 | 343 | 0 | 0 | 27 |
| 9 | szybkikotlet.pl | 50.00 | 12.27 | 2254 | 15 | 0 | 0 | 165 | 56 | 0 | 0 | 1 |
| 10^ | czyszczeniedywanowzawiercie.pl | 35.00 | 8.59 | 2304 | 15 | 0 | 0 | 126 | 129 | 0 | 0 | 5 |

^ domain name at which a website is available, * prices as of 1 March 2017, ** prices as of 1 March 2017 in USD at the average exchange rate of 1 USD = PLN 4.0756

Web site applications: 1) http://wycena-www.pl, 2) http://www.wycenastron.pl, 3) http://www.speedtest.pl/wycena, 4) http://wycen-strone.pl, 5) http://www.siteprice.org, 6) https://www.yourwebsitevalue.com, 7) http://www.siteworthtraffic.com, 8) http://www.howmuchismysite.com, 9) free-of-charge by smaller, additional websites, new page views are generated on the main website, e.g. the Speed Test website offers web applications, such as a test of internet connection's speed, various rankings and a mechanism for website appraisal. On the other hand, statistics regarding the use of the website are used as a bargaining chip for acquiring advertisers. Such applications are also used to save cookie files on users' devices, which are then used for re-marketing.

Web address appraisals are significantly different from website appraisals. This is confirmed by the results of performed tests. Domain names connected with a particular website are appraised at a higher value than a domain name on its own. In such case an expert appraisal must consider the value of the website available at a particular web address. Obviously, there is a possibility to appraise a domain name on its own on the free market, regardless of whether it is tied to a particular website. However, one should remember that an analysis of website's history constitutes a component of a domain name appraisal, including websites which have been published at a particular web address.

Appraisal of domain names and websites should be conducted only automatically and using only one tool. It should be comprehensive by involving multiple criteria and dimensions.

At present, there is no generally available web application which would perform credible domain name appraisals in an algorithmic and automated manner. It is impossible to obtain a credible website appraisal using the tested web applications. Entities which are interested in a credible verification of the value of a website or a domain name should use a paid appraisal service in the form of an appraisal report which includes a list of applied appraisal methods.